\documentclass[11pt,a4paper,useAMS]{emulateapj}

\usepackage{natbib}
\usepackage{amsmath}
\usepackage{graphicx}
\usepackage{morefloats}

\begin{document}


\title{The discovery of a radio halo in PLCK G147.3--16.6 at $z=0.65$}


\author{R.~J.~van~Weeren\altaffilmark{1,$\star$},  H.~T.~Intema\altaffilmark{2},  D.~V.~Lal\altaffilmark{3}, A.~Bonafede\altaffilmark{4},  C.~Jones\altaffilmark{1},  W.~R.~Forman\altaffilmark{1}, H.~J.~A.~R\"ottgering\altaffilmark{5}, M.~Br\"uggen\altaffilmark{4}, A.~Stroe\altaffilmark{5}, M.~Hoeft\altaffilmark{6}, S.~E.~Nuza\altaffilmark{7}, and F.~de~Gasperin\altaffilmark{4}\vspace{3mm}
}


\affil{\altaffilmark{1}Harvard-Smithsonian Center for Astrophysics, 60 Garden Street, Cambridge, MA 02138, USA}
\affil{\altaffilmark{2}National Radio Astronomy Observatory, 1003 Lopezville Road, Socorro, NM 87801-0387, USA}
\affil{\altaffilmark{3}National Centre for Radio Astrophysics, TIFR, Pune University Campus, Post Bag 3, Pune 411 007, India}

\affil{\altaffilmark{4}Hamburger Sternwarte, Gojenbergsweg 112, 21029 Hamburg, Germany}
\affil{\altaffilmark{5}Leiden Observatory, Leiden University, P.O. Box 9513, NL-2300 RA Leiden, The Netherlands}
\affil{\altaffilmark{6}Th\"uringer Landessternwarte Tautenburg, Sternwarte 5, 07778, Tautenburg, Germany}
\affil{\altaffilmark{7} Leibniz-Institut f\"ur Astrophysik Potsdam (AIP), An der Sternwarte 16, 14482 Potsdam, Germany}
\email{E-mail: rvanweeren@cfa.harvard.edu}

\altaffiltext{$\star$}{Einstein Fellow}

\shorttitle{A radio halo in PLCK G147.3--16.6}
\shortauthors{van Weeren et al.}

\vspace{0.5cm}
\begin{abstract}
\noindent 
Recent X-ray and Sunyaev-Zel'dovich (SZ) observations have dramatically increased the number of known distant galaxy clusters.  In some merging, low-redshift ($z<0.4$) clusters, centrally-located, diffuse, extended radio emission  (called radio halos) has been found. Using the Giant Metrewave Radio Telescope (GMRT), we report the detection of diffuse radio emission in the binary-merging cluster PLCK G147.3--16.6 located at $z = 0.65$. 
We classify the emission as a giant radio halo due to the large physical extent of about 0.9~Mpc and low-surface brightness. We measure an integrated flux density of $7.3\pm1.1$~mJy at 610~MHz for the radio halo, resulting in a 1.4~GHz radio power of $5.1~\times 10^{24}$ W~Hz$^{-1}$. The radio halo power is consistent with that expected from the known correlation between X-ray luminosity or the cluster integrated SZ signal and radio power. Our observations also suggest that more of these distant radio halos could be discovered with the GMRT.
\vspace{4mm}
\end{abstract}



\section{Introduction}
In some merging galaxy clusters, Mpc-scale diffuse radio synchrotron emission is found, which indicates the presence of relativistic particles and magnetic fields in the intracluster medium (ICM). The diffuse emission is classified depending on the location, morphology, and polarization properties \citep[see][for a recent review]{2012A&ARv..20...54F}.

Radio halos have sizes of about a Mpc and are centrally located in clusters \citep[e.g.][]{2011MNRAS.412....2B}. 
 It has been proposed that radio halos trace particles re-accelerated by turbulence generated during a cluster merger event \citep[][]{2001MNRAS.320..365B, 2001ApJ...557..560P}. In an alternative model, radio halos are produced by secondary electrons injected during proton-proton collisions in the ICM \citep[e.g.,][]{1980ApJ...239L..93D, 1999APh....12..169B, 2000A&A...362..151D, 2001ApJ...562..233M, 2010ApJ...722..737K}. {Secondary models are  challenged by the large energy content of cosmic ray protons needed to explain radio halos with very steep spectra \citep[e.g.,][]{2008Natur.455..944B} and by the non-detection of $\gamma$-rays from radio halos \citep[e.g.,][]{2011ApJ...728...53J,2012MNRAS.426..956B}.}

{For radio halos there is a correlation between X-ray luminosity and radio power/luminosity (the $L_{\rm{X}}-P$ correlation), with the most luminous clusters hosting the most powerful radio halos \citep[e.g.,][]{2000ApJ...544..686L,2002A&A...396...83E,2006MNRAS.369.1577C}.  There is also a strong bimodality observed in the $L_{\rm{X}}-P$ plane. Merging clusters that host radio halos follow the $L_{\rm{X}}-P$ scaling relation, while for other relaxed clusters the upper limits on the radio power are well below the correlation \citep{2007ApJ...670L...5B}. The radio halo power also correlates with the integrated Sunyaev-Zel'dovich (SZ) effect signal  \citep{2012MNRAS.421L.112B, 2013AN....334..350B, 2013ApJ...777..141C} given by the Compton $Y_{\rm{SZ}}$ parameter, which traces the integrated pressure along the line of sight. The SZ signal is expected to be a better indicator of cluster mass than the X-ray luminosity and should be less affected by the cluster dynamical state.  In addition, there are indications that the fraction of clusters with diffuse radio emission is larger for SZ selected samples \citep{2013arXiv1307.3049S}, which suggests that X-rays and SZ statistics provide complementary information on the radio-merger connection. }

 \begin{figure}[h!]
\begin{center}
\includegraphics[angle =90, trim =0cm 0cm 0cm 0cm,width=0.47\textwidth]{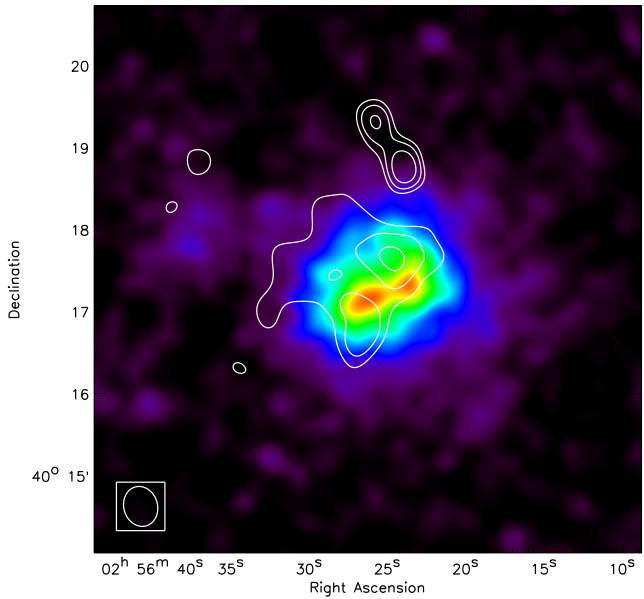}
\end{center}\vspace{-3mm}
\caption{XMM-Newton 0.5--3.0~keV X-ray image of PLCK~G147.3--16.6 convolved with a 16\arcsec~FWHM Gaussian. Radio contours from the low-resolution GMRT 610~MHz image (Fig.~\ref{fig:radiolrandhr}; right panel) are overlaid in white. Contour levels are drawn at $\sqrt{[1, 2, 4, 8, \ldots]} \times 3\sigma_{\mathrm{rms}}$, with $\sigma_{\mathrm{rms}}=0.175$~mJy~beam$^{-1}$.}
\label{fig:xray}
\end{figure}

 \begin{figure*}[ht!]
\begin{center}
\includegraphics[angle =90, trim =0cm 0cm 0cm 0cm,width=0.49\textwidth]{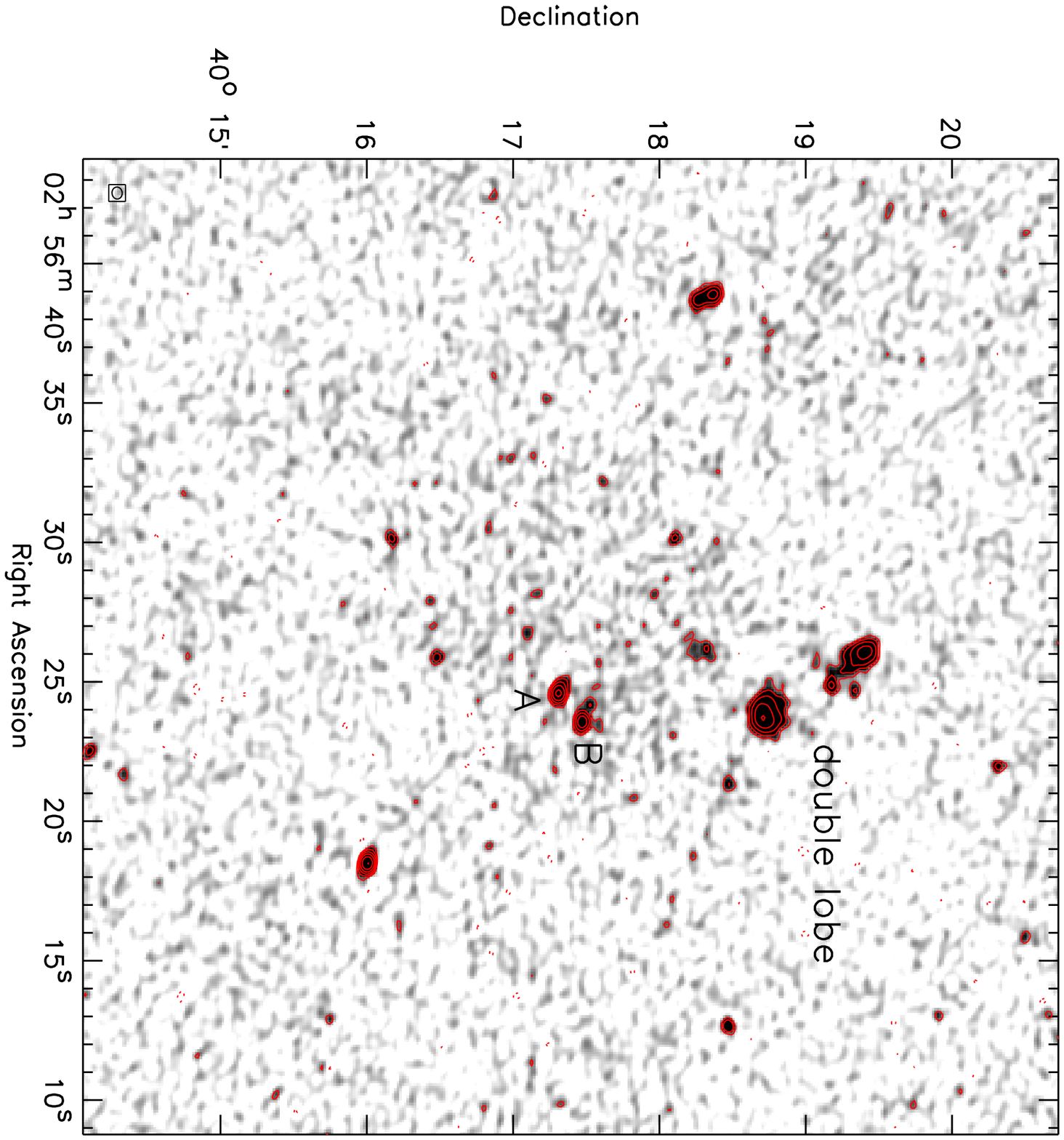}
\includegraphics[angle =90, trim =0cm 0cm 0cm 0cm,width=0.49\textwidth]{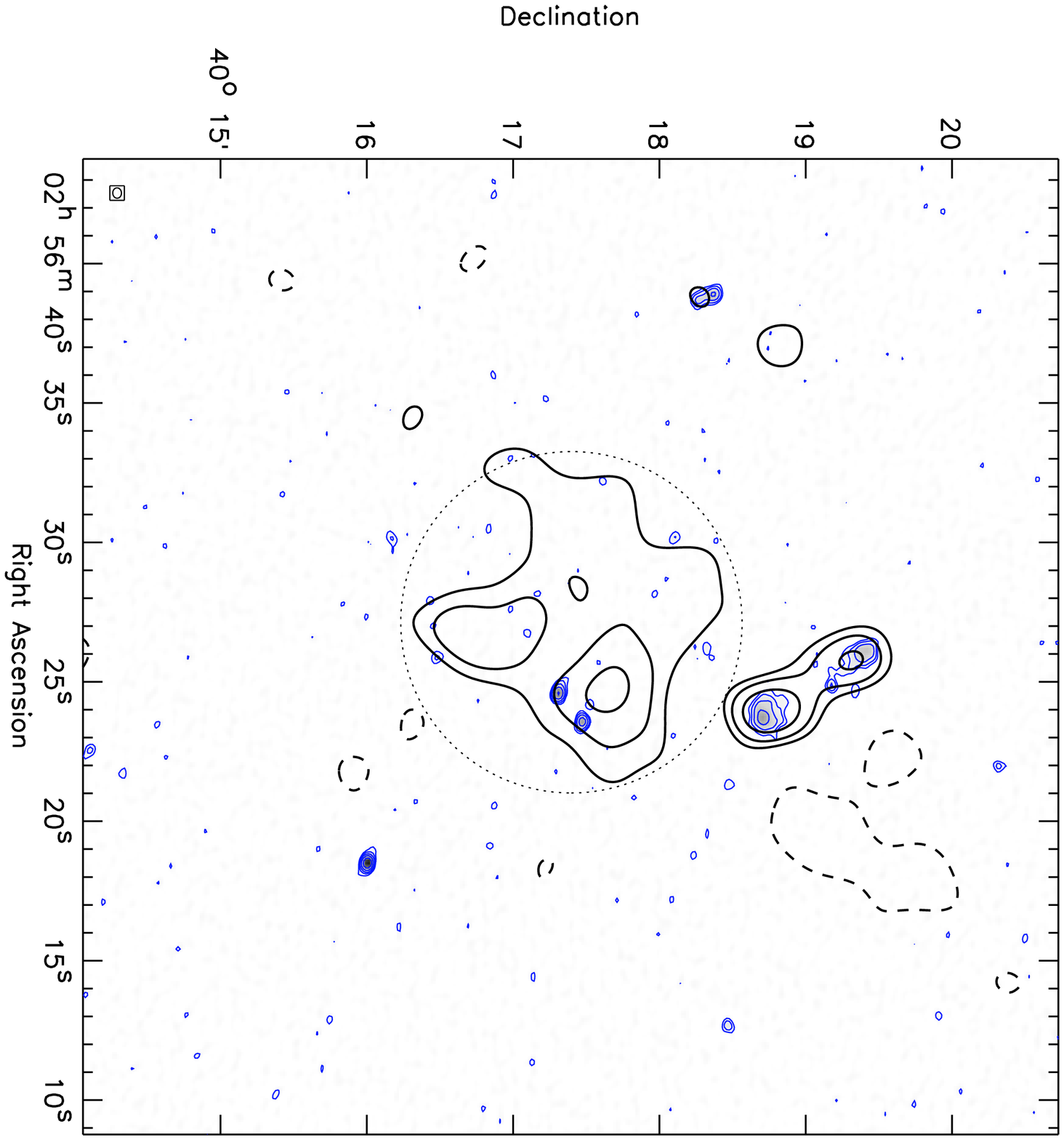}
\end{center}\vspace{-3mm}
\caption{Left: GMRT 610~MHz image (using all data). Contour levels are drawn at ${[-1, 1, 2, 4, 8, \ldots]} \times 3\sigma_{\mathrm{rms}}$, with $\sigma_{\mathrm{rms}}=38$~$\mu$Jy~beam$^{-1}$. Negative contours are shown by the dotted lines. Source A and B, both cluster members, are indicated. Right: A low-resolution $30\arcsec\times25\arcsec$ image of the cluster with the emission from compact sources subtracted. Contours are drawn at$ \sqrt{[1, 2, 4, 8, \ldots]} \times 3\sigma_{\mathrm{rms}}$, with the $2.5\sigma$ negative contours show by the dashed lines and $\sigma_{\mathrm{rms}}=0.175$~mJy~beam$^{-1}$. The high-resolution image ($4.3\arcsec\times3.6\arcsec$, indicated in the bottom left corner), made by excluding data below 2~k$\lambda$, is shown with blue contours. Contours levels are drawn at ${[1, 2, 4, 8, \ldots]} \times 3\sigma_{\mathrm{rms}}$ ($\sigma_{\mathrm{rms}}=43$~$\mu$Jy~beam$^{-1}$), this emission was subtracted before making the low-resolution image. The dotted circle indicates the area for which we performed the integrated flux measurement.}
\label{fig:radiolrandhr}
\end{figure*}

The majority of the presently known halos and relics are located at low and moderate redshifts ($z\lesssim 0.4$). 
Not much is known about the radio relic and halo population beyond $z \gtrsim 0.6$, with only a one cluster known that hosts diffuse radio emission at $z=0.87$ \citep{2012ApJ...748....7M,2013arXiv1310.6786L} and two around a redshift of 0.55 \citep{2003MNRAS.339..913E, 2009A&A...505..991V, 2009A&A...503..707B, 2012MNRAS.426...40B}. Radio halos and relics are in general difficult to study owing to their low surface brightness and steep radio spectrum, making them faint at high frequencies. Since they are already difficult to detect in nearby clusters, only the most luminous ones can be detected beyond $z \gtrsim 0.6$. According to the $L_{\rm{X}}-P$ correlation the most luminous halos are located in the most massive (i.e., X-ray luminous) clusters. We have therefore started to observe a sample of massive $z \gtrsim 0.6$ clusters with the Giant Metrewave Radio Telescope (GMRT) to carry out a first systematic characterization of the distant halo (and relic) population. In this {letter} we report the discovery of diffuse emission in the merging cluster PLCK G147.3--16.6 at $z=0.65$.

Throughout this {letter} we assume a $\Lambda$CDM cosmology with $H_{0} = 71$~km~s$^{-1}$~Mpc$^{-1}$, $\Omega_{m} = 0.27$, and $\Omega_{\Lambda} = 0.73$. With the adopted cosmology, 1\arcsec~corresponds to a physical scale of 6.9~kpc at $z=0.65$ \citep{2006PASP..118.1711W}. All images are in the J2000 coordinate system.

\section{Observations \& data reduction}
\label{sec:obs}
PLCK G147.3--16.6  was observed for 8~hrs with the GMRT on Jan. 18, 2013 at 610~MHz. Both the RR and LL correlations were recorded. We used an integration time of 8~s and, after splitting of noisy edge channels, obtained 29~MHz of usable bandwidth. The data were reduced using a semi-automatic pipeline employing the Astronomical Image Processing System\footnote{http://www.aips.nrao.edu} (AIPS), ParselTongue and Obit \citep{2008PASP..120..439C}. The reduction consisted of automatic RFI removal and subtraction \citep{2009ApJ...696..885A}, bandpass and gain calibration, and subsequent cycles of self-calibration. In addition, direction dependent phase solutions were obtained to employ the Source Peeling and Atmospheric Modeling ionospheric calibration scheme \citep{2009A&A...501.1185I}. Sources with a signal to noise ratio $> 1000$ had their own individual direction dependent gain solutions applied. The flux-scale for the primary calibrator 3C147 was set according to the \cite{2012MNRAS.423L..30S} flux scale. For the imaging, facets were used to correct for the non-coplanar nature of the array and to apply the direction dependent calibration solutions.

We used archival {\it i}-band images taken with the DOLORES (Device Optimized for the LOw RESolution) instrument on the Telescopio Nazionale Galileo (TNG) telescope at La Palma. The $2048\times 2048$ pixel images have a scale of 0.252\arcsec per pixel. We combined 10 images with a total exposure time of 3000~s using IRAF. We used a 100 pixel medium filter to self-flatten the images as no {\it i}-band flat field images were found for the night of the observing run. 
XMM-Newton archival observations (ObsID: 0679181301) of the cluster were reduced with the Science Analysis Software\footnote{http://xmm.esa.int/sas/} (SAS, v13.0), resulting in a net clean exposure time of about 10.2~ks. Periods with high background counts were filtered using the tasks \emph{pn-filter} and \emph{mos-filter}. We made an exposure-corrected  MOS+pn image in the 0.5--3.0~keV band, see Fig.~\ref{fig:xray}.

\section{PLCK G147.3--16.6}
PLCK G147.3--16.6 was confirmed to be a cluster by the \cite{2013A&A...550A.130P}. Based on XMM-Newton spectra, a redshift of 0.62  was found using the Fe~K line. The cluster has an X-ray luminosity of  $7.15\pm0.66 \times 10^{44}$~erg~$^{-1}$ within $R_{500}$ in the 0.1--2.4 keV band. Spectroscopic observations obtained with Gemini North GMOS-N of about a dozen candidate cluster members resulted in a redshift of $0.66 \pm 0.05$ for the cluster. Excluding two galaxies at $z=0.68$, the \cite{2013A&A...550A.130P} give a redshift of $0.645 \pm 0.005$. Their optical imaging shows a complex galaxy distribution, without a single dominant cD galaxy. The XMM observations reveal two peaks/substructures in the X-ray surface brightness separated by $\sim250$~kpc (see Fig.~\ref{fig:xray}), which roughly coincide with the observed galaxy distribution. There are hints that the X-ray surface brightness distribution is more complex as the eastern substructure is more elongated than the western one. The \cite{2013A&A...550A.130P} reported a global temperature ($T_{500}$) of $8.8\pm0.8$~keV using the XMM data. Both the optical and X-ray data indicate that PLCK G147.3--16.6 is undergoing a merger event. A summary of the cluster's properties is given in Table~\ref{tab:properties}.

\section{Results}
\label{sec:results}

\begin{table}[h!]
\begin{center}
\caption{Cluster and radio halo properties}
\begin{tabular}{ll}
\hline
\hline
$z$ &$0.645 \pm 0.005^{a}$\\
$R_{500}$ (kpc) & 1042\\
$L_{\rm{X, 500}}$ ($10^{44}$ erg~s$^{-1}$, 0.1--2.4~keV) & $7.15\pm 0.66$ \\ 
$Y_{500}$ ($10^{-4}$ arcmin$^{2}$) &  $5.2\pm1.7^{b}$ \\
$T_{500}$ (keV) & $8.8\pm0.8$ \\
$M_{500}$ ($10^{14}$  M$_{\odot}$ )& $6.3\pm0.4$  \\
$S_{\rm{halo,610MHz}}$  (mJy) & $7.3\pm1.1$ \\
$P_{\rm{halo,1.4GHz}}$ ($10^{24}$ W Hz$^{-1}$)& $5.1 \pm  0.8^{c}$\\
$R_{\rm{halo}}$ (Mpc) & 0.9 \\
Compact source fluxes (mJy) & $2.42 \pm 0.25$ (A)  \\
& $1.63\pm0.18$ (B) \\
&   $0.88\pm0.29$ (C)\\
&  $0.38\pm0.09$ (D) \\
& $0.31\pm0.08$ (E) \\
& $0.19\pm0.07$ (F)  \\
\hline
\hline
\end{tabular}
\label{tab:properties}
\end{center}
$^{a}$ see the \cite{2013A&A...550A.130P} for a discussion on the redshift\\
$^{b}$ $0.89\pm 0.29 \times 10^{-4}$~Mpc$^{2}$ \\
$^{c}$ assuming a spectral index of $\alpha=-1.3$ for the radio halo, with $F_{\nu} \propto \nu^{\alpha}$\\
\end{table}

\begin{figure}[h]
\begin{center}
\includegraphics[angle =90, trim =0cm 0cm 0cm 0cm,width=0.47\textwidth]{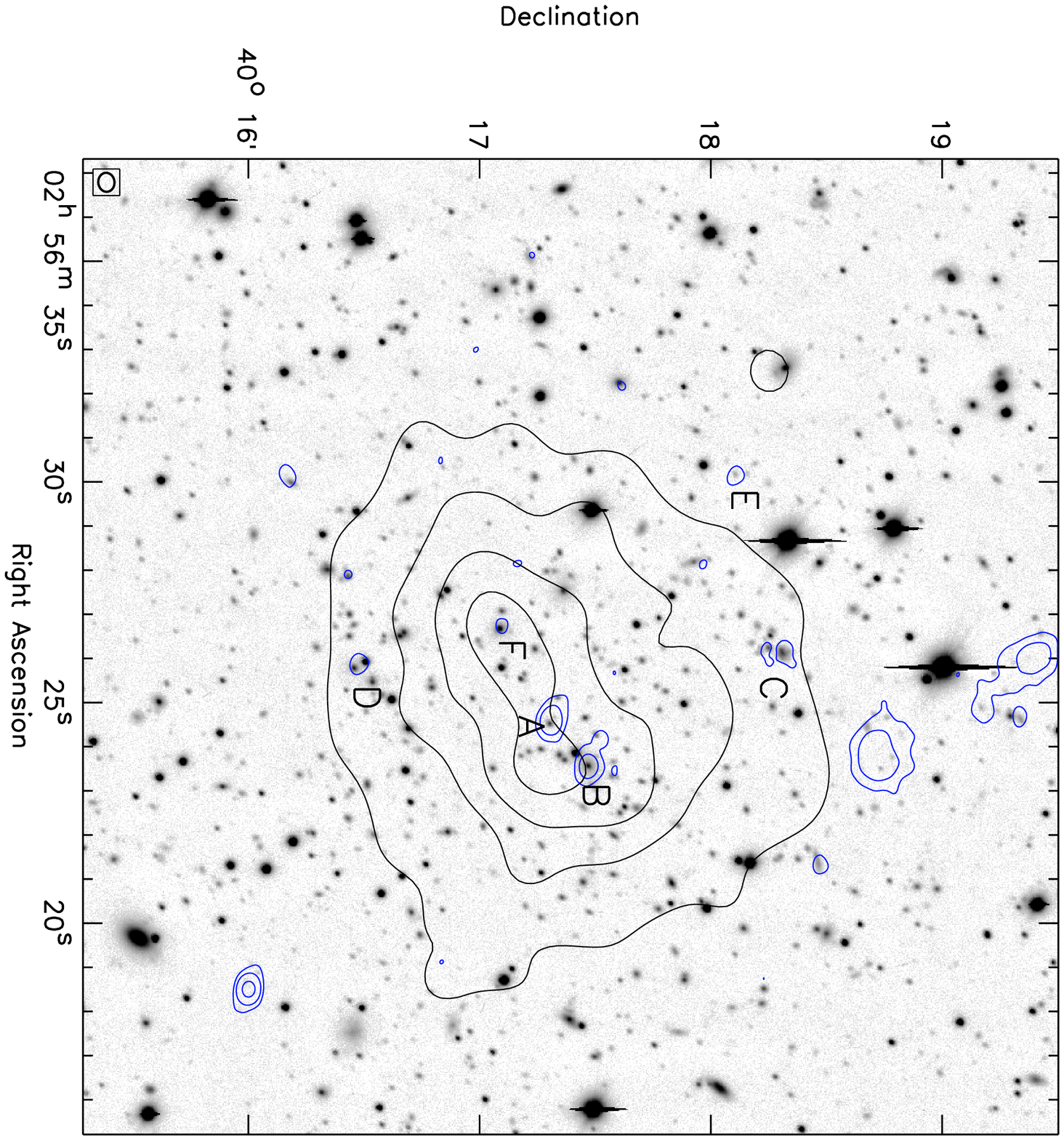}
\end{center}\vspace{-3mm}
\caption{TNG/DOLORES {\it i}-band image of PLCK~G147.3--16.6. XMM-Newton X-ray contours from Fig.~\ref{fig:xray} are overlaid in black. Radio contours from the high-resolution GMRT 610~MHz image (Fig.~\ref{fig:radiolrandhr}; left panel) are overlaid in blue. Compact sources are labelled. Contour levels are drawn at ${[1, 4, 16, 64, \ldots]} \times 4\sigma_{\mathrm{rms}}$. The beam size is $4.9\arcsec \times 4.2\arcsec$ and indicated in the bottom left corner.}
\label{fig:optical}
\end{figure}

Our GMRT 610~MHz image of the cluster is shown in Fig.~\ref{fig:radiolrandhr} (left panel). This image was made with Briggs weighting \citep{briggs_phd} using a robust value of $-1$, giving a resolution of $4.9\arcsec \times 4.2\arcsec$. Several compact radio sources are visible as well as a double lobe source just to the north of the cluster, with the two lobes separated by about 1\arcmin. Two {relatively bright} compact sources are located in the central region of the cluster. We identify optical counterpart for these sources in the TNG {\it i}-band image, see Fig.~\ref{fig:optical}. Source A is a spectroscopically confirmed cluster member. Source B is associated with the cD galaxy of the western subcluster. This galaxy is a cluster member given the very similar color to the other clusters members in fig.~3 from the \cite{2013A&A...550A.130P}. {Besides A and B, a few other fainter sources are present (see Fig.~\ref{fig:optical}). The flux densities of these sources are reported in Table~\ref{tab:properties}.}

 \begin{figure*}[ht!]
\begin{center}
\includegraphics[angle =90, trim =0cm 0cm 0cm 0cm,width=0.477\textwidth]{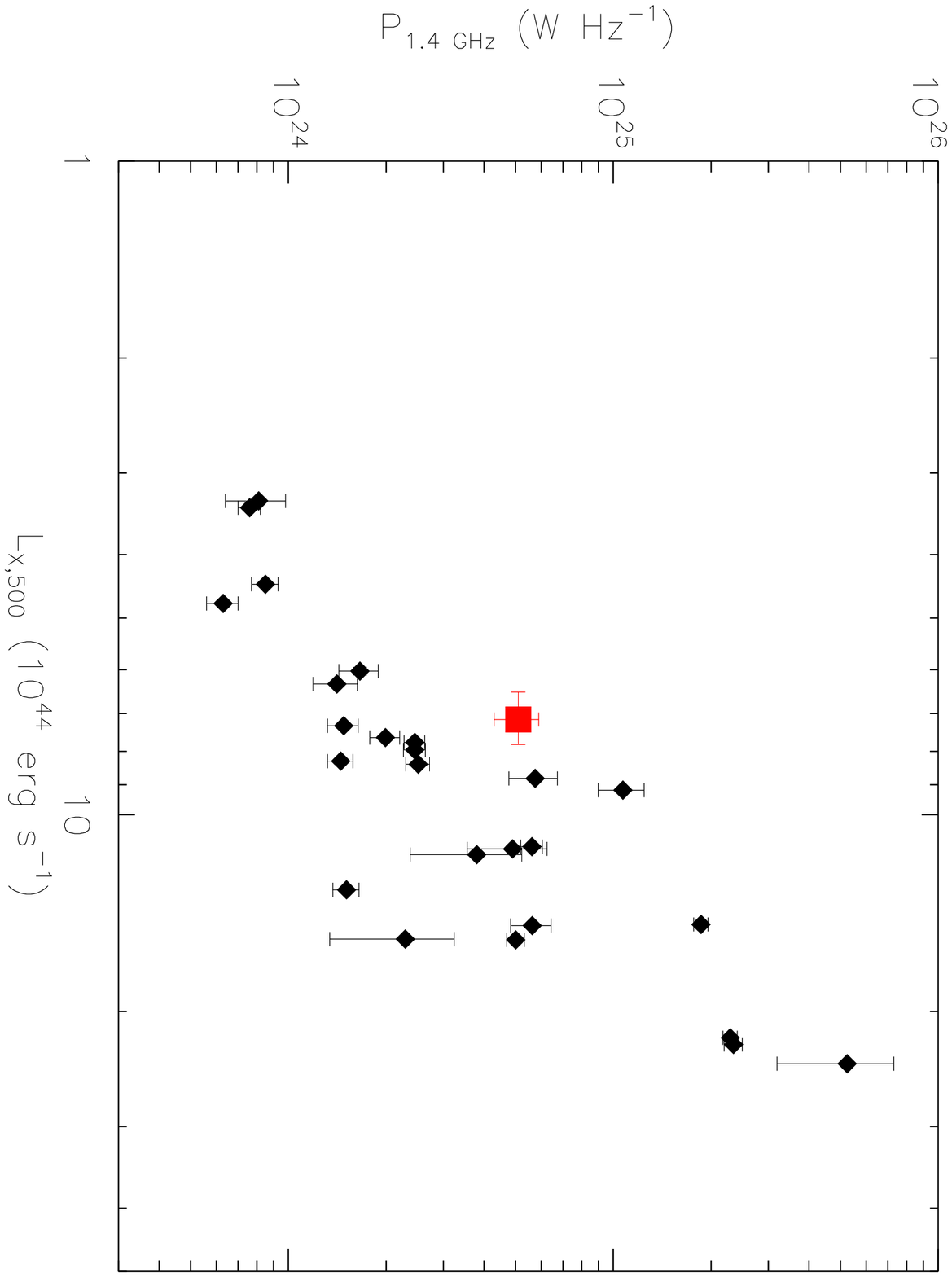}
\includegraphics[angle =90, trim =0cm 0cm 0cm 0cm,width=0.49\textwidth]{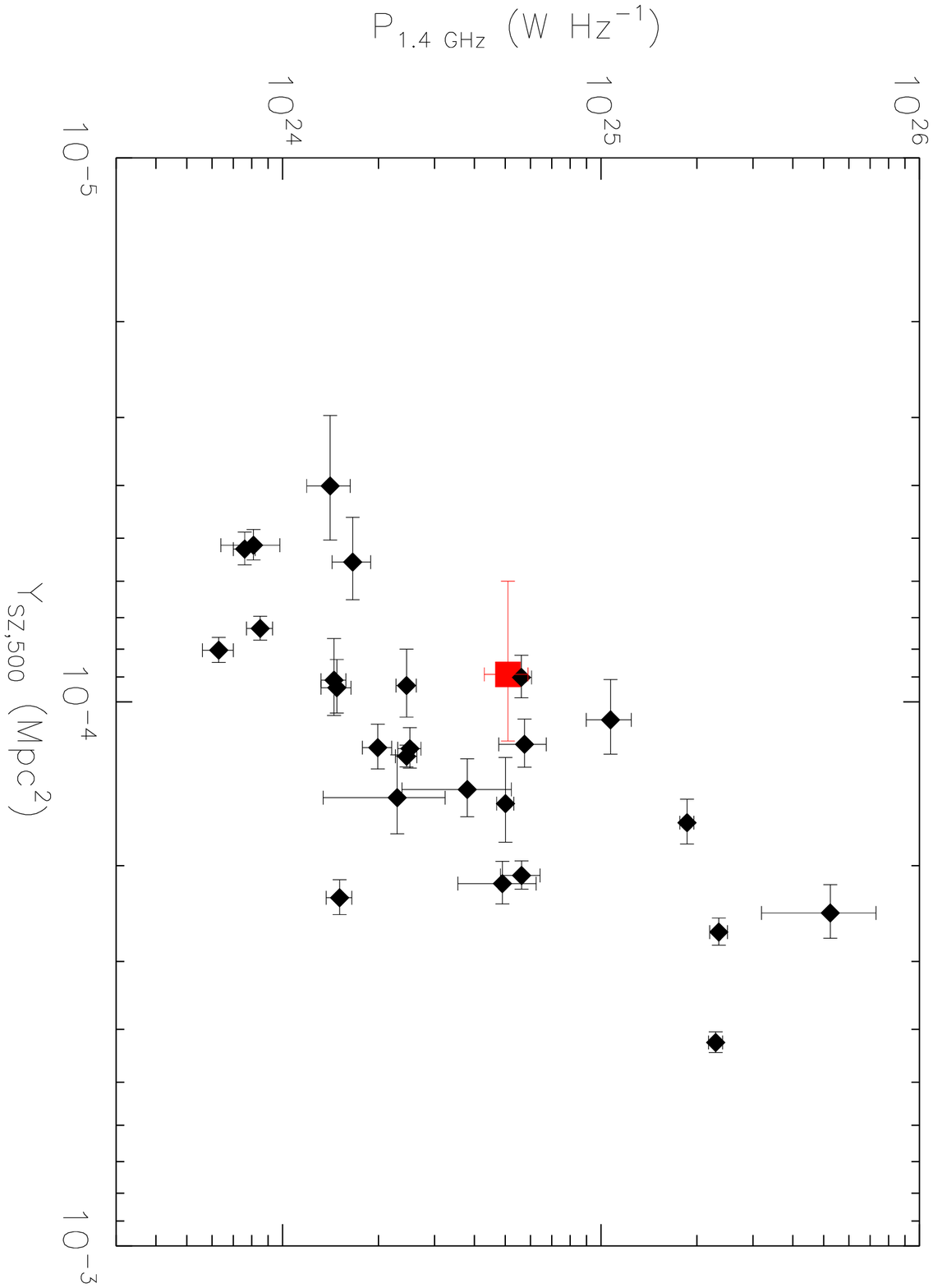}
\end{center}\vspace{-3mm}
\caption{Left:  Distribution of clusters in the $L_{\rm{X,500}} - P_{\rm{1.4~GHz}}$ plane.  The data were taken from the list compiled by \cite{2013ApJ...777..141C} and are plotted with black diamonds. The datapoint for PLCK~G147.3--16.6 is indicated by a red square.
Right: Distribution of clusters in the  $Y_{\rm{SZ,500}} - P_{\rm{1.4~GHz}}$ plane. Labeling is the same as in the left panel. We conclude that the PLCK~G147.3--16.6 radio halo power is what is expected from known $L_{\rm{X,500}} - P_{\rm{1.4~GHz}}$ and $Y_{\rm{SZ,500}} - P_{\rm{1.4~GHz}}$ correlations.\vspace{9mm}}
\label{fig:corrs}
\end{figure*}

A careful visual inspection of the image reveals an increase in the ``noise'' at the cluster position,  hinting at a large extended low-surface brightness component. To make this emission more visible, we subtracted the emission from compact radio sources from the uv-data. For this, we first made a 
 high-resolution image by excluding data below 2~k$\lambda$ (corresponding to a angular scale of about 2\arcmin). This image is shown in Fig.~\ref{fig:radiolrandhr} (right panel). We subtracted the clean components obtained from this image from the uv-data. 
 We then re-imaged the data, excluding data above 8~k$\lambda$ and using a taper of 5~k$\lambda$ (units are  the distance to the 30\% percent point of the Gaussian in k$\lambda$) to enhance diffuse emission.  This resulted in an image with a resolution of $30\arcsec\times25\arcsec$. {We verified  that the  source subtraction procedure did not leave any positive residuals, by visually inspecting the low-resolution image at the locations of several dozen compact sources that are located throughout the primary beam. } We  clearly detect diffuse emission at the cluster position with a total extent of about 2.2\arcmin, corresponding to a physical extent ($R_{\rm{halo}}$) of 0.9~Mpc at the distance of the cluster. 
 Some emission from the double lobe source to the north remains visible because of the extended nature of this source. {Source C and F were not cleaned in the data with the short uv-spacings removed (Fig.~\ref{fig:radiolrandhr}; right panel), as their peak fluxes fall below the clean box threshold. After subtracting their contribution, we measure an integrated flux ($S_{\rm{610MHz}}$) of $7.3\pm1.1$~mJy for the diffuse emission, summing the emission in the circular region indicated in Fig.~\ref{fig:radiolrandhr}. The quoted uncertainty is based on a 10\% absolute flux calibration uncertainty and the statistical uncertainty due to map noise, i.e., $(\sigma_{S_{\rm{610 MHz}}})^2 = (0.1S_{\rm{610MHz}})^2 + (\sigma_{\rm{rms}}\sqrt{\rm{\#beams\,in\,halo\,area}})^2$. We note that the actual uncertainty could be somewhat larger, as it is not fully clear what the area is over which we should integrate the halo flux.}

In Fig.~\ref{fig:xray} we overlay the radio emission on an XMM MOS+pn 0.5--3.0~keV X-ray image. The radio emission roughly follows the X-ray emission, although the radio emission seems to be somewhat more extended in the northeast direction. It should be noted however that the radio emission has a very low surface brightness and therefore not too many conclusions should be drawn based on the brightness distribution. We classify the source as a radio halo, since the emission coincides with most of the X-ray emission from the ICM. In addition, the very low surface brightness is  typical for a radio halo. The fact that the cluster is undergoing a merger, based on the optical and X-ray data, provides further support for this interpretation.

If we assume a spectral index of $-1.3$ for the radio halo {\citep[e.g.,][]{2012A&ARv..20...54F}},  we obtain a 1.4~GHz {k-corrected} radio power of  $P_{\rm{1.4~GHz}} = 5.1 \times 10^{24}$~W~Hz$^{-1}$. It is thought that distant radio halos have on average steeper spectral indices than their low-z counterparts \citep[e.g.,][]{2010A&A...509A..68C}. If we assume a spectral index of $-1.8$ this would result in a {(k-corrected)} radio power of $4.3 \times 10^{24}$~W~Hz$^{-1}$. This difference in the radio power between $\alpha=-1.3$ and $-1.8$ is thus about the same as the uncertainty introduced by the flux measurement error. The position of PLCK~G147.3--16.6 in the $L_{X,500}-P_{\rm{1.4\mbox{ }GHz}}$ and $Y_{SZ,500}-P_{\rm{1.4\mbox{ }GHz}}$ diagrams is shown in Fig.~\ref{fig:corrs}. {Based on the $L_{X}-P_{\rm{1.4\mbox{ }GHz}}$ correlation we would expect a power of about $2.3 \times 10^{24}$~W~Hz$^{-1}$, using the  \emph{BCES Bisector (RH only)} fit from \citet[][table~3]{2013ApJ...777..141C}. Given the intrinsic scatter in the correlation, measurement uncertainties and unknown spectral index, the observed radio power of PLCK~G147.3--16.6 is consistent with being on the  $L_{X}-P_{\rm{1.4\mbox{ }GHz}}$ correlation.}
Based on the $Y_{\rm{SZ,500}}-P_{\rm{1.4\mbox{ }GHz}}$ \emph{BCES Bisector (RH only)} fit from \cite{2013ApJ...777..141C}, we expect a radio power of $2.4 \times 10^{24}$~W~Hz$^{-1}$, almost the same as the prediction from the $L_{X}-P_{\rm{1.4\mbox{ }GHz}}$ relation. Therefore, we conclude that the measured radio power is in agreement with that expected from the  $L_{X}-P_{\rm{1.4\mbox{ }GHz}}$  and $Y_{SZ}-P_{\rm{1.4\mbox{ }GHz}}$ correlations.

\section{Conclusions}
\label{sec:conclusions}
We have discovered diffuse radio emission in the cluster PLCK~G147.3--16.6 with the GMRT at 610~MHz. We classify this diffuse emission as a 0.9~Mpc radio halo. For the radio halo we measure an integrated flux of $7.3\pm1.1$~mJy, resulting in a radio power of  $P_{\rm{1.4~GHz}} = 5.1\times 10^{24}$~W~Hz$^{-1}$ at the redshift of 0.65. This value is in agreement with what is expected  from the known correlations between X-ray luminosity and SZ signal and radio power. 

The detection of radio halos beyond $z\gtrsim 0.6$ has generally been considered more difficult because of the strong increase of the inverse Compton (IC) losses with redshift, which decreases the radio halo power. In addition, distant halos are expected to have steeper spectral indices than their low-z counterparts \citep[e.g.,][]{2010A&A...509A..68C}. This is caused by the increase  of the IC losses with redshift, which limit the maximum energy of the accelerated electrons in these systems, and the assumption the radio halo spectra have convex shapes as stochastic particle acceleration by MHD turbulence is believed to be an inefficient process.
The detection of a radio halo in PLCK~G147.3--16.6 at $z=0.65$ now opens up the exciting opportunity to study a radio halo in a distant cluster in more detail. Deep follow-up studies with the Karl G. Jansky VLA and LOFAR will be pursued to determine the spectral index and polarization properties of the radio halo. Our observations also suggest that detection and initial studies of distant radio halos are within the capabilities of the GMRT and that we should expect additional detections from our larger sample.

\acknowledgments
{\it Acknowledgments:}

We thank the staff of the GMRT who have made these 
observations possible. The GMRT is run by the National 
Centre for Radio Astrophysics of the Tata Institute of 
Fundamental Research. 

We used observations made with the Italian Telescopio Nazionale Galileo (TNG) operated on the island of La Palma by the Fundaci\'on Galileo Galilei of the INAF (Istituto Nazionale di Astrofisica) at the Spanish Observatorio del Roque de los Muchachos of the Instituto de Astrofisica de Canarias. This work is based on observations obtained with XMM-Newton, an ESA science mission with instruments and contributions directly funded by ESA Member States and the USA (NASA).

RJvW is supported by NASA through the Einstein Postdoctoral
grant number PF2-130104 awarded by the Chandra X-ray Center, which is
operated by the Smithsonian Astrophysical Observatory for NASA under
contract NAS8-03060. HTI acknowledges support from the National Radio Astronomy Observatory, which is a facility of the National Science Foundation operated under cooperative agreement by Associated Universities, Inc. A.B., M.B and M.H. acknowledge support by the research group FOR 1254 funded by the Deutsche Forschungsgemeinschaft: ``Magnetisation of interstellar and intergalactic media:the prospects of low-frequency radio observations''. SEN is supported by the DFG grant MU1020~16-1.



{\it Facilities:} \facility{GMRT}, \facility{XMM}, \facility{TNG (DOLORES)}


\begin{thebibliography}{34}
\expandafter\ifx\csname natexlab\endcsname\relax\def\natexlab#1{#1}\fi

\bibitem[{{Athreya}(2009)}]{2009ApJ...696..885A}
{Athreya}, R. 2009, \apj, 696, 885

\bibitem[{{Basu}(2012)}]{2012MNRAS.421L.112B}
{Basu}, K. 2012, \mnras, 421, L112

\bibitem[{{Basu}(2013)}]{2013AN....334..350B}
---. 2013, Astronomische Nachrichten, 334, 350

\bibitem[{{Blasi} \& {Colafrancesco}(1999)}]{1999APh....12..169B}
{Blasi}, P., \& {Colafrancesco}, S. 1999, Astroparticle Physics, 12, 169

\bibitem[{{Bonafede} {et~al.}(2009){Bonafede}, {Feretti}, {Giovannini},
  {Govoni}, {Murgia}, {Taylor}, {Ebeling}, {Allen}, {Gentile}, \&
  {Pihlstr{\"o}m}}]{2009A&A...503..707B}
{Bonafede}, A., {Feretti}, L., {Giovannini}, G., {et~al.} 2009, \aap, 503, 707

\bibitem[{{Bonafede} {et~al.}(2012){Bonafede}, {Br{\"u}ggen}, {van Weeren},
  {Vazza}, {Giovannini}, {Ebeling}, {Edge}, {Hoeft}, \&
  {Klein}}]{2012MNRAS.426...40B}
{Bonafede}, A., {Br{\"u}ggen}, M., {van Weeren}, R., {et~al.} 2012, \mnras,
  426, 40

\bibitem[{{Briggs}(1995)}]{briggs_phd}
{Briggs}, D.~S. 1995, PhD thesis, New Mexico Institute of Mining Technology,
  Socorro, New Mexico, USA

\bibitem[{{Brown} \& {Rudnick}(2011)}]{2011MNRAS.412....2B}
{Brown}, S., \& {Rudnick}, L. 2011, \mnras, 412, 2

\bibitem[{{Brunetti} {et~al.}(2008){Brunetti}, {Giacintucci}, {Cassano},
  {Lane}, {Dallacasa}, {Venturi}, {Kassim}, {Setti}, {Cotton}, \&
  {Markevitch}}]{2008Natur.455..944B}
{Brunetti}, G., {Giacintucci}, S., {Cassano}, R., {et~al.} 2008, \nat, 455, 944

\bibitem[{{Brunetti} {et~al.}(2012){Brunetti}, {Blasi}, {Reimer}, {Rudnick},
  {Bonafede}, \& {Brown}}]{2012MNRAS.426..956B}
{Brunetti}, G., {Blasi}, P., {Reimer}, O., {et~al.} 2012, \mnras, 426, 956



\bibitem[{{Brunetti} {et~al.}(2001){Brunetti}, {Setti}, {Feretti}, \&
  {Giovannini}}]{2001MNRAS.320..365B}
{Brunetti}, G., {Setti}, G., {Feretti}, L., \& {Giovannini}, G. 2001, \mnras,
  320, 365

\bibitem[{{Brunetti} {et~al.}(2007){Brunetti}, {Venturi}, {Dallacasa},
  {Cassano}, {Dolag}, {Giacintucci}, \& {Setti}}]{2007ApJ...670L...5B}
{Brunetti}, G., {Venturi}, T., {Dallacasa}, D., {et~al.} 2007, \apjl, 670, L5

\bibitem[{{Cassano} {et~al.}(2010){Cassano}, {Brunetti}, {R{\"o}ttgering}, \&
  {Br{\"u}ggen}}]{2010A&A...509A..68C}
{Cassano}, R., {Brunetti}, G., {R{\"o}ttgering}, H.~J.~A., \& {Br{\"u}ggen}, M.
  2010, \aap, 509, A68+

\bibitem[{{Cassano} {et~al.}(2006){Cassano}, {Brunetti}, \&
  {Setti}}]{2006MNRAS.369.1577C}
{Cassano}, R., {Brunetti}, G., \& {Setti}, G. 2006, \mnras, 369, 1577

\bibitem[{{Cassano} {et~al.}(2013){Cassano}, {Ettori}, {Brunetti},
  {Giacintucci}, {Pratt}, {Venturi}, {Kale}, {Dolag}, \&
  {Markevitch}}]{2013ApJ...777..141C}
{Cassano}, R., {Ettori}, S., {Brunetti}, G., {et~al.} 2013, \apj, 777, 141




\bibitem[{{Cotton}(2008)}]{2008PASP..120..439C}
{Cotton}, W.~D. 2008, \pasp, 120, 439

\bibitem[{{Dennison}(1980)}]{1980ApJ...239L..93D}
{Dennison}, B. 1980, \apjl, 239, L93

\bibitem[{{Dolag} \& {En{\ss}lin}(2000)}]{2000A&A...362..151D}
{Dolag}, K., \& {En{\ss}lin}, T.~A. 2000, \aap, 362, 151

\bibitem[{{Edge} {et~al.}(2003){Edge}, {Ebeling}, {Bremer}, {R{\"o}ttgering},
  {van Haarlem}, {Rengelink}, \& {Courtney}}]{2003MNRAS.339..913E}
{Edge}, A.~C., {Ebeling}, H., {Bremer}, M., {et~al.} 2003, \mnras, 339, 913

\bibitem[{{En{\ss}lin} \& {R{\"o}ttgering}(2002)}]{2002A&A...396...83E}
{En{\ss}lin}, T.~A., \& {R{\"o}ttgering}, H. 2002, \aap, 396, 83

\bibitem[{{Feretti} {et~al.}(2012){Feretti}, {Giovannini}, {Govoni}, \&
  {Murgia}}]{2012A&ARv..20...54F}
{Feretti}, L., {Giovannini}, G., {Govoni}, F., \& {Murgia}, M. 2012, \aapr, 20,
  54


\bibitem[{{Intema} {et~al.}(2009){Intema}, {van der Tol}, {Cotton}, {Cohen},
  {van Bemmel}, \& {R{\"o}ttgering}}]{2009A&A...501.1185I}
{Intema}, H.~T., {van der Tol}, S., {Cotton}, W.~D., {et~al.} 2009, \aap, 501,
  1185

\bibitem[{{Jeltema} \& {Profumo}(2011)}]{2011ApJ...728...53J}
{Jeltema}, T.~E. \& {Profumo}, S. 2011, \apj, 728, 53




\bibitem[{{Keshet} \& {Loeb}(2010)}]{2010ApJ...722..737K}
{Keshet}, U., \& {Loeb}, A. 2010, \apj, 722, 737

\bibitem[{{Liang} {et~al.}(2000){Liang}, {Hunstead}, {Birkinshaw}, \&
  {Andreani}}]{2000ApJ...544..686L}
{Liang}, H., {Hunstead}, R.~W., {Birkinshaw}, M., \& {Andreani}, P. 2000, \apj,
  544, 686

\bibitem[{{Lindner} {et~al.}(2013){Lindner}, {Baker}, {Hughes}, {Battaglia},
  {Gupta}, {Knowles}, {Marriage}, {Menanteau}, {Moodley}, {Reese}, \&
  {Srianand}}]{2013arXiv1310.6786L}
{Lindner}, R.~R., {Baker}, A.~J., {Hughes}, J.~P., {et~al.} 2013, ArXiv
  e-prints

\bibitem[{{Menanteau} {et~al.}(2012){Menanteau}, {Hughes}, {Sif{\'o}n},
  {Hilton}, {Gonz{\'a}lez}, {Infante}, {Barrientos}, {Baker}, {Bond}, {Das},
  {Devlin}, {Dunkley}, {Hajian}, {Hincks}, {Kosowsky}, {Marsden}, {Marriage},
  {Moodley}, {Niemack}, {Nolta}, {Page}, {Reese}, {Sehgal}, {Sievers},
  {Spergel}, {Staggs}, \& {Wollack}}]{2012ApJ...748....7M}
{Menanteau}, F., {Hughes}, J.~P., {Sif{\'o}n}, C., {et~al.} 2012, \apj, 748, 7

\bibitem[{{Miniati} {et~al.}(2001){Miniati}, {Jones}, {Kang}, \&
  {Ryu}}]{2001ApJ...562..233M}
{Miniati}, F., {Jones}, T.~W., {Kang}, H., \& {Ryu}, D. 2001, \apj, 562, 233

\bibitem[{{Petrosian}(2001)}]{2001ApJ...557..560P}
{Petrosian}, V. 2001, \apj, 557, 560

\bibitem[{{Planck Collaboration} {et~al.}(2013){Planck Collaboration}, {Ade},
  {Aghanim}, {Arnaud}, {Ashdown}, {Aumont}, {Baccigalupi}, {Balbi}, {Banday},
  {Barreiro}, {Bartlett}, {Battaner}, {Benabed}, {Beno{\^i}t}, {Bernard},
  {Bersanelli}, {Bikmaev}, {B{\"o}hringer}, {Bonaldi}, {Bond}, {Borgani},
  {Borrill}, {Bouchet}, {Brown}, {Burigana}, {Butler}, {Cabella}, {Carvalho},
  {Catalano}, {Cay{\'o}n}, {Chamballu}, {Chary}, {Chiang}, {Chon},
  {Christensen}, {Clements}, {Colafrancesco}, {Colombi}, {Coulais}, {Crill},
  {Cuttaia}, {Da Silva}, {Dahle}, {Davis}, {de Bernardis}, {de Gasperis}, {de
  Zotti}, {Delabrouille}, {D{\'e}mocl{\`e}s}, {D{\'e}sert}, {Diego}, {Dolag},
  {Dole}, {Donzelli}, {Dor{\'e}}, {Douspis}, {Dupac}, {En{\ss}lin}, {Eriksen},
  {Finelli}, {Flores-Cacho}, {Forni}, {Frailis}, {Franceschi}, {Frommert},
  {Galeotta}, {Ganga}, {G{\'e}nova-Santos}, {Giraud-H{\'e}raud},
  {Gonz{\'a}lez-Nuevo}, {Gonz{\'a}lez-Riestra}, {G{\'o}rski}, {Gregorio},
  {Gruppuso}, {Hansen}, {Harrison}, {Hempel}, {Henrot-Versill{\'e}},
  {Hern{\'a}ndez-Monteagudo}, {Herranz}, {Hildebrandt}, {Hivon}, {Hobson},
  {Holmes}, {Hornstrup}, {Hovest}, {Huffenberger}, {Hurier}, {Jaffe},
  {Jagemann}, {Jones}, {Juvela}, {Kneissl}, {Knoche}, {Knox}, {Kunz},
  {Kurki-Suonio}, {Lagache}, {Lamarre}, {Lasenby}, {Lawrence}, {Le Jeune},
  {Leach}, {Leonardi}, {Liddle}, {Lilje}, {Linden-V{\o}rnle},
  {L{\'o}pez-Caniego}, {Luzzi}, {Mac{\'{\i}}as-P{\'e}rez}, {Maino},
  {Mandolesi}, {Mann}, {Maris}, {Marleau}, {Marshall},
  {Mart{\'{\i}}nez-Gonz{\'a}lez}, {Masi}, {Massardi}, {Matarrese}, {Mazzotta},
  {Mei}, {Meinhold}, {Melchiorri}, {Melin}, {Mendes}, {Mennella}, {Mitra},
  {Miville-Desch{\^e}nes}, {Moneti}, {Morgante}, {Mortlock}, {Munshi},
  {Naselsky}, {Nati}, {Natoli}, {N{\o}rgaard-Nielsen}, {Noviello}, {Osborne},
  {Pajot}, {Paoletti}, {Perdereau}, {Perrotta}, {Piacentini}, {Piat},
  {Pierpaoli}, {Piffaretti}, {Plaszczynski}, {Platania}, {Pointecouteau},
  {Polenta}, {Popa}, {Poutanen}, {Pratt}, {Prunet}, {Puget}, {Reinecke},
  {Remazeilles}, {Renault}, {Ricciardi}, {Rocha}, {Rosset}, {Rossetti},
  {Rubi{\~n}o-Mart{\'{\i}}n}, {Rusholme}, {Sandri}, {Savini}, {Scott}, {Smoot},
  {Stanford}, {Stivoli}, {Sudiwala}, {Sunyaev}, {Sutton}, {Suur-Uski},
  {Sygnet}, {Tauber}, {Terenzi}, {Toffolatti}, {Tomasi}, {Tristram},
  {Valenziano}, {Van Tent}, {Vielva}, {Villa}, {Vittorio}, {Wade}, {Wandelt},
  {Welikala}, {Weller}, {White}, {Yvon}, {Zacchei}, \&
  {Zonca}}]{2013A&A...550A.130P}
{Planck Collaboration}, {Ade}, P.~A.~R., {Aghanim}, N., {et~al.} 2013, \aap,
  550, A130

\bibitem[{{Scaife} \& {Heald}(2012)}]{2012MNRAS.423L..30S}
{Scaife}, A.~M.~M., \& {Heald}, G.~H. 2012, \mnras, 423, L30

\bibitem[{{Sommer} \& {Basu}(2013)}]{2013arXiv1307.3049S}
{Sommer}, M.~W., \& {Basu}, K. 2013, ArXiv e-prints

\bibitem[{{van Weeren} {et~al.}(2009){van Weeren}, {R{\"o}ttgering},
  {Br{\"u}ggen}, \& {Cohen}}]{2009A&A...505..991V}
{van Weeren}, R.~J., {R{\"o}ttgering}, H.~J.~A., {Br{\"u}ggen}, M., \& {Cohen},
  A. 2009, \aap, 505, 991

\bibitem[{{Wright}(2006)}]{2006PASP..118.1711W}
{Wright}, E.~L. 2006, \pasp, 118, 1711

\end{thebibliography}
\end{document}